\documentclass[conference]{IEEEtran}
\IEEEoverridecommandlockouts
\usepackage{cite}
\usepackage{amsmath,amssymb,amsfonts}
\usepackage{algorithm}
\usepackage{algpseudocode}
\usepackage{graphicx}
\usepackage{textcomp}
\usepackage{xcolor}
\usepackage{url}

\def\BibTeX{{\rm B\kern-.05em{\sc i\kern-.025em b}\kern-.08em
    T\kern-.1667em\lower.7ex\hbox{E}\kern-.125emX}}
\begin{document}

\title{Physics-Guided Bayesian Optimization for High-Dimensional Mixed-Variable\\MIMO Base Station Design
\thanks{This work was supported by JST PRESTO under Grant Number JPMJPR23P3, CRONOS under Grant Number JPMJCS24N1, BOOST under Grant Number JPMJBS2415, and JSPS KAKENHI under Grant Number 25K00138.}
}

\author{\IEEEauthorblockN{Koki~Kanzaki~and~Koya~Sato}%
\IEEEauthorblockA{Artificial Intelligence eXploration Research Center,\\
The University of Electro-Communications, Chofu, Tokyo, Japan\\
Email: k-kanzaki@uec.ac.jp, k\_sato@ieee.org
}
}

\maketitle

\begin{abstract}

This paper proposes a physics-guided Bayesian optimization for high-dimensional mixed-variable multiple-input and multiple-output (MIMO) base station (BS) design. The considered problem jointly selects a subset of candidate sites for BS deployment and optimizes the azimuth angles, downtilt angles, and transmit power spectral densities of the BSs, while each configuration is evaluated using computationally expensive site-specific ray tracing.
To efficiently optimize the system configuration, the proposed method constructs a low-cost physics-based proxy from precomputed propagation information. The proxy-estimated communication coverage is used as the Gaussian process (GP) prior mean, and a residual GP with three-dimensional physical features learns the discrepancy between the proxy and full evaluations. Ray-tracing-based evaluations in two urban scenarios show that the proposed method achieves up to approximately 15 percentage points higher coverage than conventional and high-dimensional optimization baselines under the same evaluation budget.
\end{abstract}

\begin{IEEEkeywords}
Bayesian optimization, Gaussian process, base station configuration, ray tracing, MIMO
\end{IEEEkeywords}

\section{Introduction}
Wireless infrastructure is increasingly deployed and configured for specific environments, as exemplified by private 5G\,\cite{wenPrivate5GNetworks2022} and enterprise Wi-Fi networks\,\cite{geraciWiFi25Years2025}. In parallel, detailed open three-dimensional environment models are becoming widely available, including Project PLATEAU in Japan\footnote{\url{https://www.mlit.go.jp/plateau/en/}} and Helsinki 3D\footnote{\url{https://www.hel.fi/en/decision-making/information-on-helsinki/maps-and-geospatial-data/helsinki-3d}}. Together with site-specific propagation simulators such as Sionna RT\,\cite{sionna}, these resources have made it increasingly feasible to predict wireless channels before physical deployment. These developments create an opportunity to design base-station (BS) infrastructure directly against the communication performance expected in the actual deployment environment, rather than relying primarily on simplified propagation models or coverage heuristics. Such site-specific design must jointly account for BS locations, antenna orientations, and transmit powers, because blockage, reflection, and diffraction from surrounding structures strongly affect communication performance.
This problem has become more important as modern wireless systems increasingly rely on multi-antenna transmission with channel-adaptive precoding and beamforming. Even when BSs operate independently without coordinated transmission, the communication performance of a deployment is governed by location-dependent spatial multiple-input multiple-output (MIMO) channels and the resulting inter-BS interference. Therefore, received power or static antenna patterns alone are insufficient to characterize the quality of a BS configuration. High-fidelity channel simulation, such as ray tracing over a three-dimensional environment model, provides a natural means of evaluating such configurations. However, repeatedly performing site-specific ray tracing over many receiver locations inside a joint placement-and-configuration search is computationally expensive\,\cite{vasudevanMachineLearningRadio2024}.
An efficient optimization method is therefore needed to identify effective BS configurations with only a limited number of ray-tracing evaluations. \par
Bayesian optimization (BO)\,\cite{snoekPracticalBayesianOptimization2012a} is well suited to this setting and has been applied to BS placement optimization\,\cite{satoBayesianOptimizationFramework2024a}.
However, many existing studies on BS placement optimization assume that BSs can be deployed at arbitrary locations within a prescribed area.
In practice, deployment constraints, such as the availability of suitable installation space, often prevent BSs from being placed at arbitrary locations, restricting deployment to a finite set of feasible candidate sites.
The resulting BS design problem is a conditional mixed-variable optimization problem that combines discrete variables for selecting candidate sites for BS deployment with continuous parameters, such as transmit power and antenna orientation, that are relevant only to the BSs deployment at the selected candidate sites.
Moreover, the dimensionality of the design problem increases with the number of candidate sites, making it difficult to handle these variables directly using a naive BO implementation.
Methods for discrete optimization include BO of combinatorial structures\,\cite{baptistaBayesianOptimizationCombinatorial2018a} and combinatorial Bayesian optimization using the graph Cartesian product\,\cite{ohCombinatorialBayesianOptimization2019a}.
Trust region BO (TuRBO)\,\cite{erikssonScalableGlobalOptimization2019} and random embedding BO (REMBO)\,\cite{wangBayesianOptimizationBillion2016} have also been proposed for high-dimensional optimization.
Nevertheless, BO methods that efficiently handle high-dimensional optimization problems with conditional mixed variables, such as BS design, remain insufficiently explored. \par
Further, modern wireless systems commonly employ multi-antenna transmission with channel-adaptive precoding and beamforming.
Hence, the performance of a BS deployment depends not only on received power or static antenna patterns, but also on user-dependent spatial MIMO channels, the resulting precoders, and inter-cell interference.
This aspect is particularly important for BS design because the effective transmission pattern varies dynamically with the user location and channel state.
Nevertheless, existing BS placement formulations often rely on received-power metrics or single-stream directional transmission and do not explicitly account for channel-adaptive MIMO operation.
Therefore, BS placement and configuration should be evaluated based on the communication performance achievable over the spatial MIMO channels after precoding and interference mitigation. \par
To address these challenges, we propose a physics-guided BO for high-dimensional mixed-variable MIMO BS design.
The proposed method jointly optimizes the BS placements and their parameters; the azimuth angles, downtilt angles, and transmission powers.
The objective is to maximize coverage, defined as the fraction of target-area locations achieving a prescribed spectral efficiency, with each full evaluation based on MIMO channels obtained through site-specific ray tracing.
To efficiently search the system configuration under a limited full-evaluation budget, we construct a low-cost physics-based proxy using propagation information precomputed for the candidate sites and a discrete set of antenna orientations.
The proxy-estimated coverage is used as the prior mean of a Gaussian process (GP), while low-dimensional physical features derived from the proxy characterize the communication conditions of each configuration.
A residual GP then learns the discrepancy between the proxy and full evaluations in this physical feature space, thereby guiding the selection of subsequent configurations for expensive full evaluation.

\section{System Model}
\label{sec:system_model}
\begin{figure}[t]
    \centering
    \includegraphics[width=\linewidth]{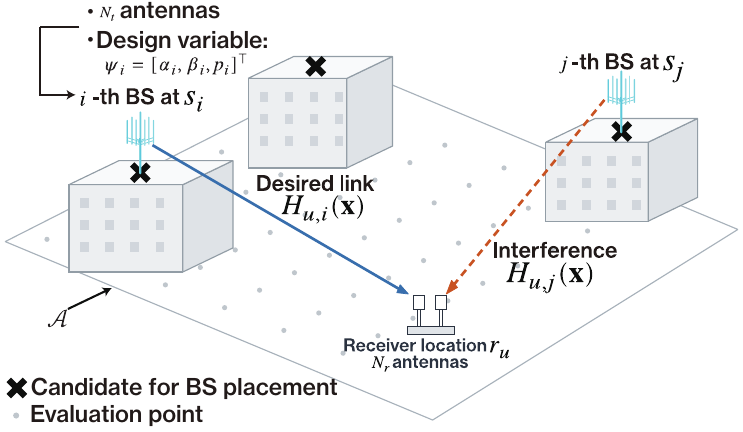}
    \caption{System model.}
    \label{fig:system_model}
\end{figure}

Fig.\,\ref{fig:system_model} summarizes the system model.
We consider a wireless system where multiple MIMO BSs operating on a shared frequency band\footnote{Owing to the orthogonality among frequency channels, the proposed method can be extended to multi-channel systems by exploiting an additive decomposition of the objective function \cite{binois2022survey}. Since this extension is straightforward, this paper focuses on system design over a shared frequency band.} are deployed in an area $\mathcal{A} \subset \mathbb{R}^3$.
Our goal is to optimize BS placements and their parameters to maximize communication area.
Possible applications of our discussion include cellular systems, private 5G, and public Wi-Fi.
\subsection{Signal Model}
Let $\mathcal{C}=\{1,\ldots,N\}$ denote the set of candidate-site
indices and $\mathbf{s}_i\in\mathbb{R}^3$ the location of candidate
site $i$.
We refer to the BS deployed at candidate site $i$ as the $i$-th BS.
A system configuration is specified by a set $\mathbf{x}\triangleq\left\{\mathcal{C}_{\mathbf{x}},\{\boldsymbol{\psi}_i\}_{i\in\mathcal{C}_{\mathbf{x}}}\right\}$,
where $\mathcal{C}_{\mathbf{x}}\subseteq\mathcal{C}$ is the set of selected candidate-site indices and $\boldsymbol{\psi}_i=[\alpha_i,\beta_i,p_i]^\top$ is the continuous parameter vector of the $i$-th BS.
Here, $\alpha_i\,\mathrm{[rad]}$, $\beta_i\,\mathrm{[rad]}$, and $p_i\,\mathrm{[W/Hz]}$ denote the azimuth angle, the downtilt angle measured from the horizontal plane, and the total transmit power spectral density (PSD) across all transmit antennas, respectively.
We evaluate the downlink spectral efficiency independently at
$N_{\mathrm{eval}}$ receiver locations
$\{\mathbf r_u\}_{u=1}^{N_{\mathrm{eval}}}\subset\mathcal A$.
These locations are evaluation points rather than simultaneously active
users. All selected BSs are assumed active; the serving BS transmits the signal to
the receiver under evaluation, while the others generate inter-BS
interference. Resource sharing and inter-user scheduling are not modeled.
\par
For simplicity, quantities associated with the $u$-th receiver location are indexed by $u$.
The channel matrix between the $i$-th BS and the $u$-th receiver location at $f_c$ is denoted by $\mathbf{H}_{u,i}(\mathbf{x}) \in \mathbb{C}^{N_r\times N_t}$, where $N_t$ and $N_r$ are the numbers of transmit and receive antenna, respectively, including polarization ports.
Each BS employs a digital MIMO architecture with an independent radio-frequency chain for each transmit antenna.
BSs operate independently, without coordinated transmission or joint precoding.
Perfect instantaneous channel state information (CSI) for the link between a receiver and its serving BS is assumed to be available at both ends and is used to precode the desired signal.\par
For a receiver at the $u$-th location served by the $i$-th BS, the singular value decomposition (SVD) of the channel matrix is
\begin{equation}
    \mathbf{H}_{u,i}(\mathbf{x}) = \mathbf{U}_{u,i}(\mathbf{x}) \boldsymbol{\Sigma}_{u,i}(\mathbf{x}) \mathbf{V}_{u,i}^{H}(\mathbf{x}).
\end{equation}
Here, $\mathbf{U}_{u,i}(\mathbf{x})\in\mathbb{C}^{N_r\times N_r}$ and $\mathbf{V}_{u,i}(\mathbf{x})\in\mathbb{C}^{N_t\times N_t}$ are unitary matrices, while $\boldsymbol{\Sigma}_{u,i}(\mathbf{x})\in\mathbb{R}_{\geq 0}^{N_r\times N_t}$ contains the singular values in nonincreasing order on its main diagonal.
The superscript $(\cdot)^H$ denotes the conjugate transpose.
The number of spatial streams, $N_s$, is fixed and satisfies $1\leq N_s\leq\min(N_t,N_r)$.
We denote by $\mathbf{V}_{u,i}^{(N_s)}(\mathbf{x}) \in \mathbb{C}^{N_t\times N_s}$ the matrix formed by the first $N_s$ columns of $\mathbf{V}_{u,i}(\mathbf{x})$, which correspond to the $N_s$ largest singular values.
The $i$-th BS allocates its total transmit PSD $p_i$ equally among the $N_s$ streams.
The transmit covariance density matrix of the desired signal is then
\begin{equation}
    \mathbf{Q}^{\mathrm{sig}}_{u,i}(\mathbf{x}) = \frac{p_i}{N_s} \mathbf{V}_{u,i}^{(N_s)}(\mathbf{x}) \left( \mathbf{V}_{u,i}^{(N_s)}(\mathbf{x}) \right)^H.
\end{equation}
The corresponding desired-signal covariance density matrix at the receiver is
\begin{equation}
    \label{eq:signal_covariance}
    \mathbf{S}_{u,i}(\mathbf{x}) = \mathbf{H}_{u,i}(\mathbf{x}) \mathbf{Q}^{\mathrm{sig}}_{u,i}(\mathbf{x}) \mathbf{H}_{u,i}^{H}(\mathbf{x}).
\end{equation}

Next, we consider interference from non-serving BSs.
For non-serving BSs, we assume isotropically distributed beam
directions independent of the channel to the receiver under evaluation.
Since $\mathbb{E}[\mathbf W_j\mathbf W_j^{\mathrm H}]
=(N_s/N_t)\mathbf I_{N_t}$, the interference-plus-noise covariance
for the $u$-th receiver served by BS $i$ is
\begin{equation}
\mathbf{\Omega}_{u|i}(\mathbf{x})
=
\sum_{\substack{j\in\mathcal C_{\mathbf{x}}\\j\neq i}}
\frac{p_j}{N_t}
\mathbf H_{u,j}(\mathbf{x})
\mathbf H_{u,j}^{\mathrm H}(\mathbf{x})
+
N_0\mathbf I_{N_r}.
\end{equation}
Assuming independent complex Gaussian signals for the desired streams and treating interference as Gaussian noise with the covariance density specified above, the spectral efficiency of the link from the $i$-th BS to the $u$-th receiver location in $\mathrm{bps/Hz}$ is
\begin{equation}
    \label{eq:link_spectral_efficiency}
    \eta_{u,i}(\mathbf{x}) = \log_2 \det \left( \mathbf{I}_{N_r} + \boldsymbol{\Omega}_{u|i}^{-1}(\mathbf{x}) \mathbf{S}_{u,i}(\mathbf{x}) \right).
\end{equation}
The receiver at the $u$-th location is associated with the BS at the selected candidate site that provides the highest reference received PSD.
We define the reference received PSD metric for the link from the $i$-th BS to the $u$-th receiver location as
\begin{equation}
    \label{eq:reference_power}
    \rho_{u,i}(\mathbf{x}) \triangleq \frac{p_i}{N_t} \operatorname{tr} \left( \mathbf{H}_{u,i}(\mathbf{x}) \mathbf{H}_{u,i}^{H}(\mathbf{x}) \right),
\end{equation}
where $\operatorname{tr}(\cdot)$ denotes the matrix trace.
The serving BS index $i_u^\ast(\mathbf{x})$ is then selected as
\begin{equation}
    \label{eq:association}
    i_u^\ast(\mathbf{x}) \in \underset{i\in\mathcal{C}_{\mathbf{x}}}{\operatorname{arg\,max}} \,\rho_{u,i}(\mathbf{x}).
\end{equation}
The spectral efficiency at the $u$-th receiver location is therefore
$R_u(\mathbf{x}) = \eta_{u,i_u^\ast(\mathbf{x})}(\mathbf{x})$.
\subsection{Problem Setting}
We define coverage as the fraction of the $N_{\mathrm{eval}}$ receiver locations satisfying the required spectral efficiency $R_{\mathrm{req}}>0$.
Specifically, the objective function is given by
\begin{equation}
    \label{eq:objective_function}
    f(\mathbf{x}) \triangleq \frac{1}{N_{\mathrm{eval}}} \sum_{u=1}^{N_{\mathrm{eval}}} \mathbf{1} \left[ R_u(\mathbf{x}) \geq R_{\mathrm{req}} \right],
\end{equation}
where $\mathbf{1}[\cdot]$ is the indicator function that outputs $1$ if the condition in brackets holds and $0$ otherwise.
Finally, the optimization problem is formulated as
\begin{equation}
    \label{eq:placement_problem}
    \begin{aligned}
        \underset{\mathbf{x}}{\operatorname{maximize}}\quad &f(\mathbf{x})\\
        \operatorname{subject\ to}\quad
        &\mathcal{C}_{\mathbf{x}}\subseteq\mathcal{C},\quad |\mathcal{C}_{\mathbf{x}}|=M,\\
        &0\leq\alpha_i<2\pi, \quad i\in\mathcal{C}_{\mathbf{x}},\\
        &\beta_{\min}\leq\beta_i\leq\beta_{\max}, \quad i\in\mathcal{C}_{\mathbf{x}},\\
        &p_{\min}\leq p_i\leq p_{\max},\quad i\in\mathcal{C}_{\mathbf{x}},
    \end{aligned}
\end{equation}
where $1\leq M\leq N$.
The constants $\beta_{\min}$ and $\beta_{\max}$ bound the downtilt angle, while constants $p_{\min}$ and $p_{\max}$ bound the transmit PSD.
We denote the set of configurations satisfying these constraints by $\mathcal{X}$.
Because radio propagation is strongly affected by surrounding buildings, evaluating $f(\mathbf{x})$ requires high-fidelity channel simulation for each configuration $\mathbf{x}$.
Moreover, since $\mathbf{x}$ contains both discrete and continuous variables, the problem in Eq.\,\eqref{eq:placement_problem} is a conditional mixed-variable black-box optimization problem that combines discrete site selection with continuous parameters defined only for the BSs deployed at the selected candidate sites.

\section{Physics-Guided Bayesian Optimization}
\label{sec:physics_guided_bayesian_optimization}
\begin{figure}
    \centering
    \includegraphics[width=\linewidth]{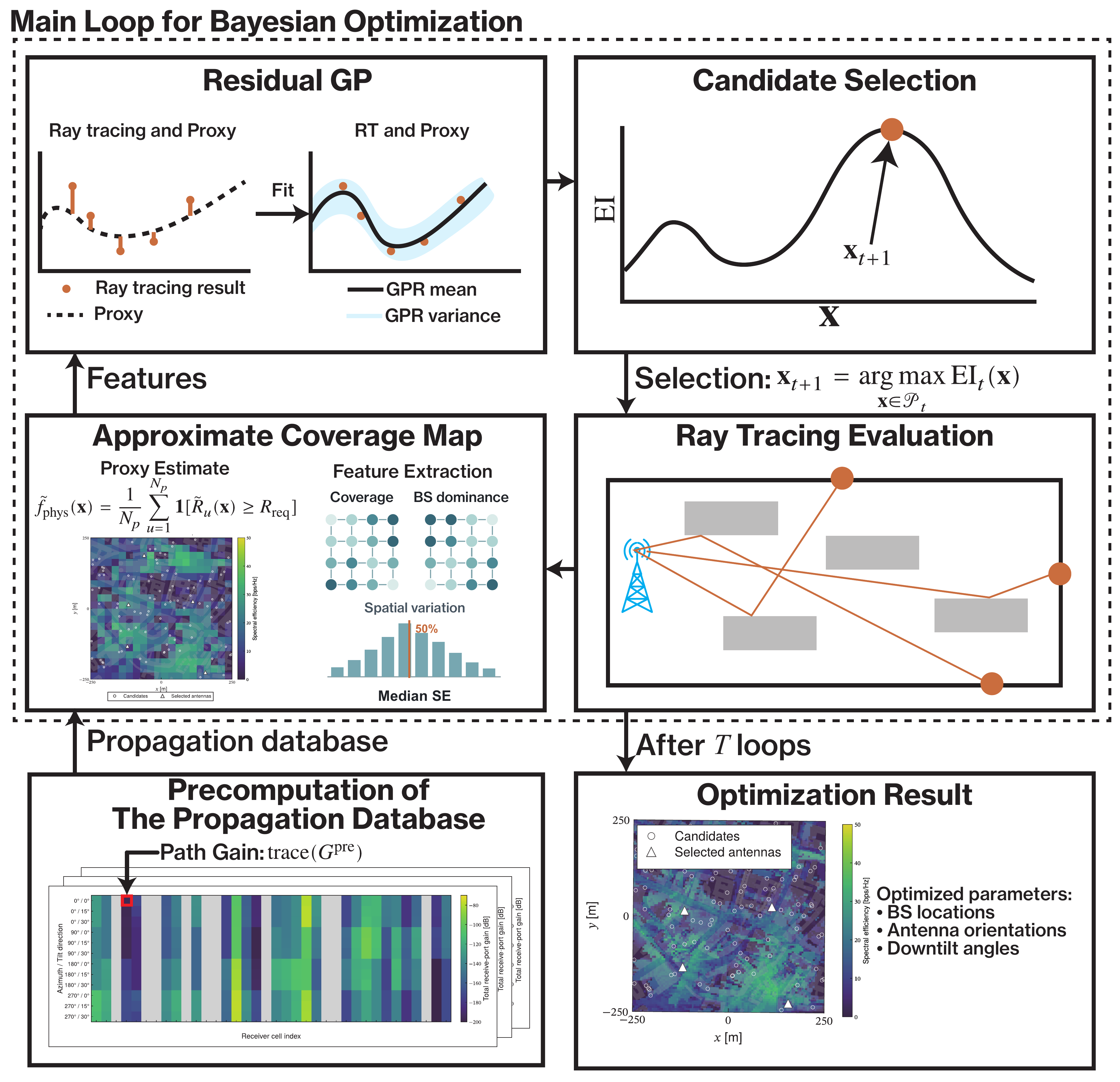}
    \vspace{-5mm}
    \caption{Overview of the proposed method.}
    \vspace{-3mm}
    \label{fig:overview_of_the_proposed_method}
\end{figure}

Since evaluating $f(\mathbf{x})$ requires costly site-specific ray
tracing, we employ BO with a physics-guided surrogate.
A GP directly defined over the variables in Eq.\,\eqref{eq:placement_problem} is problematic because
its dimension grows with the number of candidate sites and geometric
proximity does not necessarily imply similar propagation conditions.
We therefore construct a low-cost physics-based proxy and use its
low-dimensional physical features as GP inputs, as summarized in Fig.\,\ref{fig:overview_of_the_proposed_method}.

\subsection{Construction of the Physics-Based Proxy}
To avoid ray tracing for every candidate configuration, we precompute propagation information for the candidate sites, representative antenna orientations, and a reduced set of proxy receiver points before optimization begins.
We precompute propagation information for representative antenna orientations and
$N_p<N_{\mathrm{eval}}$ proxy receiver points $\{\mathbf{r}_u^p\}_{u\in\mathcal{U}_p}\subset\mathcal{A}$, where $\mathcal{U}_p=\{1,\ldots,N_p\}$.
The antenna azimuth and downtilt angles are discretized into the sets
\begin{equation}
    \label{eq:proxy_set}
    \mathcal{D}_\alpha \triangleq \{ \overline{\alpha}_1, \overline{\alpha}_2, \cdots, \overline{\alpha}_{N_\alpha} \}, \,\, \mathcal{D}_\beta \triangleq \{ \overline{\beta}_1, \overline{\beta}_2, \cdots, \overline{\beta}_{N_\beta} \},
\end{equation}
where $N_\alpha$ and $N_\beta$ are the numbers of representative azimuth and downtilt angles, respectively.
Let $\mathbf{H}^{\mathrm{pre}}_{u,i,a,b}\in\mathbb{C}^{N_r\times N_t}$ denote the MIMO channel at $f_c$ between proxy $u$-th receiver point and a BS at $i$-th candidate site with azimuth $\overline{\alpha}_a$ and downtilt $\overline{\beta}_b$.
These channel matrices are computed once by ray tracing before optimization begins.
For each channel matrix, we compute the normalized Gram matrix
\begin{equation}
    \label{eq:precomputed_gram}
    \mathbf{G}_{u,i,a,b}^{\mathrm{pre}} \triangleq \frac{1}{N_t} \mathbf{H}_{u,i,a,b}^{\mathrm{pre}} \left( \mathbf{H}_{u,i,a,b}^{\mathrm{pre}} \right)^H.
\end{equation}
We refer to the collection of these precomputed Gram matrices over all candidate sites, representative antenna orientations, and proxy receiver points as the propagation database.
Because the angles $\alpha_i$ and $\beta_i$ in a configuration $\mathbf{x}$ are continuous, the Gram matrix for an arbitrary antenna orientation is approximated by interpolating the neighboring matrices in the propagation database: we define this operation as
\begin{equation}
    \label{eq:proxy_gram}
    \tilde{\mathbf{G}}_{u,i}(\mathbf{x}) \triangleq \mathcal{I}\left( \left\{ \mathbf{G}_{u,i,a,b}^{\mathrm{pre}} \right\}; \alpha_i, \beta_i \right),
\end{equation}
where $\mathcal{I}$ denotes bilinear interpolation over neighboring azimuth and downtilt grid points, with the azimuth treated as periodic at $0$ and $2\pi$.
The interpolated matrix is a weighted sum of Gram matrices with nonnegative weights that sum to $1$, preserving their Hermitian positive semidefinite property. \par
We use the approximate Gram matrices to compute approximate spectral efficiency under the communication model in Sec.\,\ref{sec:system_model}.
Let $\tilde{\lambda}_{u,i,\ell}(\mathbf{x})$ denote the eigenvalues of $\tilde{\mathbf{G}}_{u,i}(\mathbf{x})$ in descending order and $\tilde{\mathbf{u}}_{u,i,\ell}(\mathbf{x})$ the corresponding unit-norm eigenvectors.
Accounting for the normalization by $1/N_t$ in Eq.\,\eqref{eq:precomputed_gram}, the approximate desired-signal and interference covariance density matrices are, respectively,
\begin{align}
    \label{eq:proxy_signal_covariance}
    \tilde{\mathbf{S}}_{u,i}(\mathbf{x}) &= \frac{p_iN_t}{N_s}\sum_{\ell=1}^{N_s} \tilde{\lambda}_{u,i,\ell}(\mathbf{x}) \tilde{\mathbf{u}}_{u,i,\ell}(\mathbf{x}) \tilde{\mathbf{u}}_{u,i,\ell}^{H}(\mathbf{x}), \\
    \label{eq:proxy_interference_covariance}
    \tilde{\mathbf{J}}_{u\leftarrow j}(\mathbf{x}) &= p_j\tilde{\mathbf{G}}_{u,j}(\mathbf{x}).
\end{align}
Using the approximated Gram matrix, the reference received PSD metric is approximated as $\tilde{\rho}_{u,i}(\mathbf{x}) = p_i\operatorname{tr}\left(\tilde{\mathbf{G}}_{u,i}(\mathbf{x})\right)$.
The BS with the largest value of this metric is selected as the serving BS.
The above covariance density matrices are then used to compute the approximate spectral efficiency $\tilde{R}_u(\mathbf{x})$.
The proxy estimate of the objective is defined as
\begin{equation}
    \label{eq:physics_prior_mean}
    \tilde{f}_{\mathrm{phys}}(\mathbf{x}) \triangleq \frac{1}{N_p} \sum_{u = 1}^{N_p} \mathbf{1} \left[ \tilde{R}_{u}(\mathbf{x}) \geq R_{\mathrm{req}} \right].
\end{equation}
The proxy requires no additional ray tracing for each configuration, but antenna-orientation interpolation and receiver-point subsampling introduce a discrepancy from the full objective $f(\mathbf{x})$.
We therefore use the proxy as a baseline and write the surrogate as
\begin{equation}
    \label{eq:physics_residual_model}
    g(\mathbf{x}) \triangleq \tilde{f}_{\mathrm{phys}}(\mathbf{x}) + r(\mathbf{z}(\mathbf{x})),
\end{equation}
with $f(\mathbf{x}) \approx g(\mathbf{x})$, where $\mathbf{z}(\mathbf{x})$ is defined in Sec.\,III-B and the residual function $r(\cdot)$ is learned by Gaussian process regression (GPR)\,\cite{rasmussenGaussianProcessesMachine2008a} in Sec.\,\ref{ssec:sequential_search_using_a_residial_gp}.

\subsection{Physical Feature}
The residual term in Eq.\,\eqref{eq:physics_residual_model} should reflect similarity in communication conditions rather than in raw configuration parameters.
A direct representation is $4N$-dimensional, includes continuous parameters of unselected candidate sites, and may poorly reflect propagation similarity because of blockage and reflection.
We therefore derive low-dimensional physical features from the proxy outputs.
In addition to approximate coverage, the proxy provides spatial distributions of spectral efficiency and reference received PSD.
Their local variations are relevant to receiver-grid subsampling near the spectral-efficiency threshold and BS-association changes caused by orientation interpolation when competing BSs have similar received PSDs.
We characterize these variations together with a representative spectral-efficiency level.

To quantify spatial variation, we construct an undirected graph $\mathcal{G}_p=(\mathcal{V}_p,\mathcal{E}_p)$ over the proxy receiver grid, where $\mathcal{V}_p=\mathcal{U}_p$ and $\mathcal{E}_p$ connects horizontally or vertically adjacent receiver points, with $\mathcal{E}_p\neq\emptyset$.
For scalar values $\mathbf{q}=[q_1,q_2,\ldots,q_{N_p}]^\top$ defined on the receiver points, we define the graph energy as
\begin{equation}
    \label{eq:graph_energy}
    E_{\mathcal{G}_p}(\mathbf{q})
    = \frac{1}{\lvert \mathcal{E}_p \rvert}
    \sum_{(u,v)\in\mathcal{E}_p}(q_u-q_v)^2.
\end{equation}
\par
We define a smooth measure of spectral efficiency satisfaction, $c_u(\mathbf{x})$, and a measure of serving-base-station dominance, $d_u(\mathbf{x})$, as
\begin{align}
    \label{eq:soft_coverage}
    c_u(\mathbf{x}) &\triangleq \operatorname{sigmoid}\left( \frac{\tilde{R}_u(\mathbf{x}) - R_{\mathrm{req}}}{\tau} \right), \\
    \label{eq:reference_power_gap}
    d_{u}(\mathbf{x}) &\triangleq \frac{\tilde{\rho}_{u,(1)}(\mathbf{x}) - \tilde{\rho}_{u,(2)}(\mathbf{x})}{\tilde{\rho}_{u,(1)}(\mathbf{x}) + N_r N_0},
\end{align}
where $\operatorname{sigmoid}(a)=(1+e^{-a})^{-1}$ and $\tau>0$ is a smoothing parameter with the same units as spectral efficiency.
Smaller values of $\tau$ yield a sharper transition in $c_u$ near the threshold.
The quantities $\tilde{\rho}_{u,(1)}$ and $\tilde{\rho}_{u,(2)}$ are the largest and second-largest reference received PSD metrics among the BSs deployed at the selected candidate sites.
The term $N_rN_0$ is the total noise PSD across all receive antenna.
The metric $d_u$ is small when the two strongest BSs provide similar received PSDs and large when the strongest BS dominates both the others and the noise.
Defining $\mathbf{c}(\mathbf{x})=[c_u(\mathbf{x})]_{u\in\mathcal{U}_p}$ and $\mathbf{d}(\mathbf{x})=[d_u(\mathbf{x})]_{u\in\mathcal{U}_p}$, we construct the physical feature vector $ \mathbf{h}(\mathbf{x})\in \mathbb{R}^3$ as
\begin{equation}
    \label{eq:physics_feature}
    \mathbf{h}(\mathbf{x}) \!=\!\! \left[
        E_{\mathcal{G}_p}(\mathbf{c}(\mathbf{x})),
        E_{\mathcal{G}_p}(\mathbf{d}(\mathbf{x})),
        \underset{u\in\mathcal{U}_p}{\operatorname{median}}\left(\frac{\tilde{R}_u(\mathbf{x})}{R_{\mathrm{req}}}\right)
    \right]^\top,
\end{equation}
where $\operatorname{median}(\cdot)$ denotes the median.
The first feature measures local variation in spectral efficiency satisfaction, and the second measures local variation in serving-base-station dominance.
The third expresses the representative spectral efficiency in the target area relative to the required spectral efficiency. \par
Because the features have different scales, we evaluate a set of feasible configurations using only the proxy and compute the mean and standard deviation of each feature component.
These configurations are generated independently of the initial optimization samples.
Using these statistics, we standardize the feature vector: this operation is defined as $\mathbf{z}(\mathbf{x}) \triangleq \operatorname{Standardize}(\mathbf{h}(\mathbf{x}))$.
The standardized vector is used as the GPR input.
The standardization statistics are fixed before BO begins and are computed without full objective evaluations.
Because the features depend only on communication conditions produced by the BSs deployed at the selected candidate sites, they are independent of the continuous parameters associated with unselected candidate sites.
The GPR input dimension is always 3, regardless of the number of candidate sites or deployed BSs.

\subsection{Sequential Search Using a Residual GP}
\label{ssec:sequential_search_using_a_residial_gp}
Let $\mathcal{T}_t= \{(\mathbf{x}_j,y_j) \}_{j=1}^t$ denote the observations obtained from $t$ full evaluations.
The surrogate model assumes independent observation noise $\epsilon_j\sim\mathcal{N}(0,\sigma_{\mathrm{obs}}^2)$, so that $y_j=f(\mathbf{x}_j) + \epsilon_j$.
Following the decomposition in Eq.\,\eqref{eq:physics_residual_model}, the GPR model is applied only to the discrepancy between the full evaluation and the physics-based proxy.
For observation $y_j$, the residual is defined as $r_j \triangleq y_j - \tilde{f}_{\mathrm{phys}}(\mathbf{x}_j)$.
We introduce a latent function $r:~\mathbb{R}^3 \to \mathbb{R}$ to model this residual in the physical feature space.
Before each GP fit, we compute the empirical mean $\bar{r}_t$ of the observed residuals and a positive scale $s_t$ from their empirical standard deviation.
Unlike the fixed input-feature normalization, these target statistics are updated whenever the GP is refit.
We define the standardized latent residual as $\hat{r}_t(\mathbf{z})=(r(\mathbf{z})-\bar{r}_t)/s_t$ and and model it as $\hat{r}_t(\mathbf{z}) \sim \mathrm{GP}(0, k(\mathbf{z}, \mathbf{z}'))$.
The kernel hyperparameters and the standardized observation-noise variance $\sigma_{\mathrm{obs}}^2/s_t^2$ are learned jointly whenever the GP is refit.
In standardized residual units, the covariance between configurations $\mathbf{x}$ and $\mathbf{x}'$ is $k(\mathbf{z}(\mathbf{x}),\mathbf{z}(\mathbf{x}'))$. \par
Let $\mathcal{T}_t^r=\{(\mathbf{z}(\mathbf{x}_j),r_j)\}_{j=1}^t$ denote the residual observations.
Conditioning the GP on the centered and scaled residual observations yields a posterior in standardized units.
We restore its mean by multiplying by $s_t$ and adding $\bar{r}_t$, and its variance by multiplying by $s_t^2$, obtaining the residual posterior mean $\mu_t^r(\mathbf{z})$ and variance $(\sigma_t^r(\mathbf{z}))^2$ on the original coverage scale.
Since the proxy value is deterministic for a given configuration, the predictive mean and variance of the full latent objective are
\begin{align}
    \label{eq:physics_posterior_mean}
    \mu_t^{\mathrm{phys}}(\mathbf{x}) &=\tilde{f}_{\mathrm{phys}}(\mathbf{x})+\mu_t^r(\mathbf{z}(\mathbf{x})), \\
    \label{eq:physics_posterior_variance}
    (\sigma_t^{\mathrm{phys}}(\mathbf{x}))^2 &=(\sigma_t^r(\mathbf{z}(\mathbf{x})))^2.
\end{align}
Thus, $\tilde{f}_{\mathrm{phys}}(\mathbf{x})$ acts as the physics-based prior mean, while the residual GP provides the correction and its uncertainty. \par
As the acquisition function, we use expected improvement (EI) defined as $\mathrm{EI}_t(\mathbf{x})= \mathbb{E}\left[\max(g(\mathbf{x})-f_t^+,0)\mid\mathcal{T}_t\right]$, where $f_t^+=\max_{j \leq t}y_j$ is the best observation value.
The expectation is taken under the Gaussian predictive distribution with the mean and variance given in Eqs.\,\eqref{eq:physics_posterior_mean} and \eqref{eq:physics_posterior_variance}, respectively.
This formulation compares configurations with high proxy-predicted coverage and configurations that are promising after accounting for residual corrections and uncertainty on the same objective scale. \par
To maximize the acquisition function, we generate a finite candidate set $\mathcal{P}_t$ of unevaluated configurations in the original feasible set $\mathcal{X}$.
In addition to random configurations drawn from the full search space, we generate candidates from configurations with high observed objective values in two ways: replacing one selected selected candidate site with another candidate site, or keeping the selected candidate sites fixed and modifying the continuous parameters of the BSs deployed at those sites.
All candidates satisfy the prescribed BS count and parameter bounds; duplicate and previously evaluated configurations are removed.
We compute the proxy values, physical features, and EI for this candidate set and select the next configuration as
\begin{equation}
    \label{eq:physics_next_design}
    \mathbf{x}_{t+1}\in\underset{\mathbf{x}\in\mathcal{P}_t}{\operatorname{arg\,max}} \,\mathrm{EI}_t(\mathbf{x}).
\end{equation}
Dimension reduction is applied only to the GPR input through $\mathbf{z}(\mathbf{x})$; candidate configurations are generated and selected in the original feasible set $\mathcal{X}$.
After evaluating $N_{\mathrm{init}}$ randomly selected initial configurations, we alternate GP updates and configuration selection until the evaluation budget is exhausted.
The budget $T$ counts full configuration evaluations, including the initial observations, with $1 \leq N_{\mathrm{init}} \leq T$.
Proxy and acquisition-function evaluations, as well as propagation-database precomputation, are excluded from this count.

\section{Performance Evaluation}
\begin{figure}
    \centering
    \includegraphics[width=\linewidth]{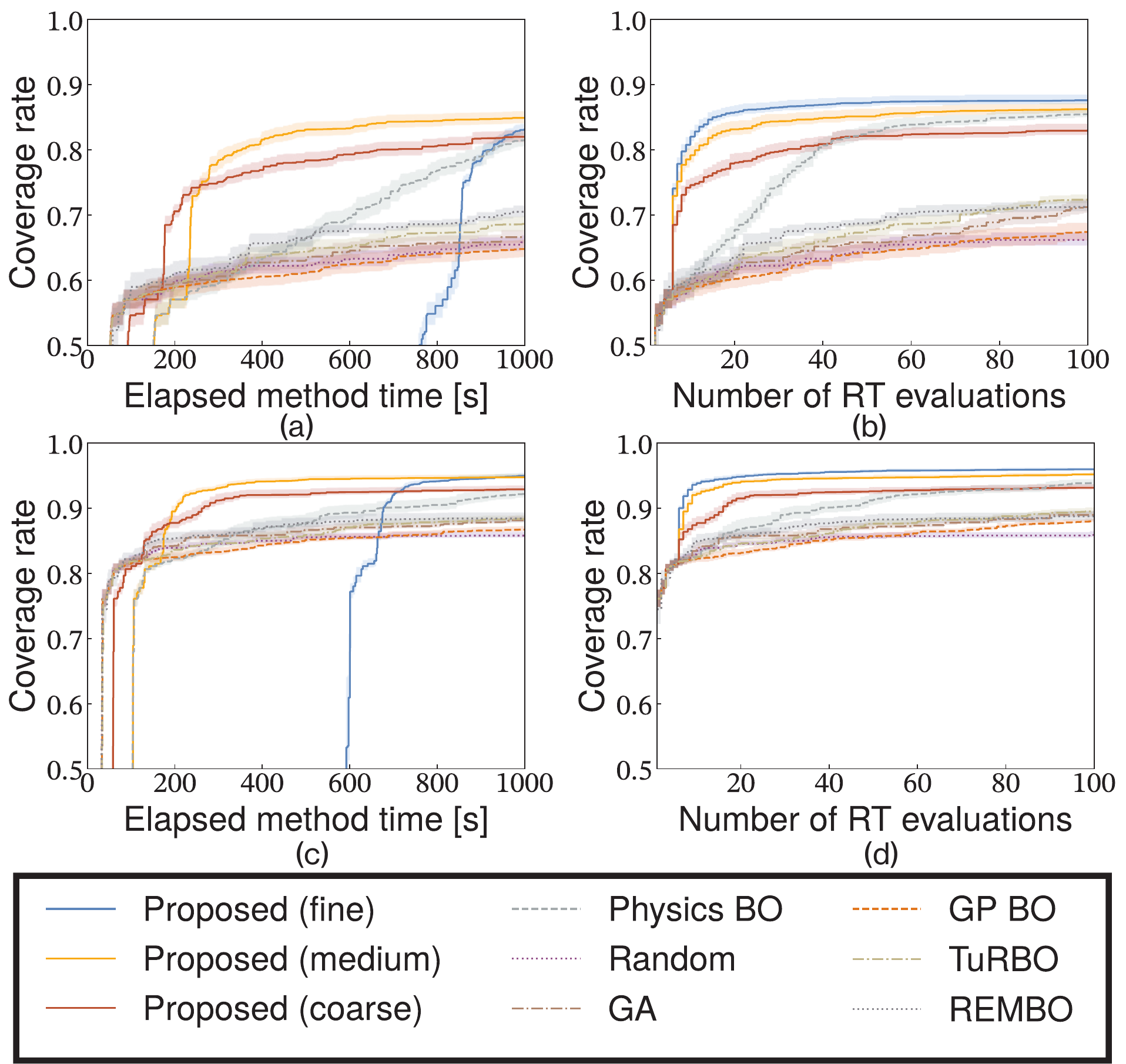}
    \caption{Best observed coverage in Munich and Etoile versus (a), (c) elapsed time and (b), (d) number of ray tracing evaluations.}
    \vspace{-3mm}
    \label{fig:result}
\end{figure}
We consider the \texttt{Munich} and \texttt{Etoile} urban scenes provided by Sionna RT\,\cite{sionna}.
We randomly place $N=100$ candidate sites $5\,\mathrm{m}$ above the rooftops and select $M=5$ of these sites for BS deployment.
The search ranges are $[0,2\pi)$ for azimuth, $[0,\pi/3]$ for downtilt, and $[-80,-60]\,\mathrm{dBm/Hz}$ for total transmit PSD.
Each transmitter uses an $8\times4$ array, and each receiver uses a $1\times2$ array.
Both arrays use crossed polarization, half-wavelength element spacing, and directional elements following 3GPP TR~38.901\,\cite{3GPP_TR38901}, giving $N_t=64$ and $N_r=4$ antenna.
The number of spatial streams is fixed at $N_s=2$.
The carrier center frequency is $f_c=3.5\,\mathrm{GHz}$.
Further, we set $N_0=4.00\times10^{-21}$ W/Hz (290 K, 0-dB noise figure).
For each full evaluation, we use a $100 \times 100$ receiver grid ($N_{\mathrm{eval}} = 10^4$) with $5\,\mathrm{m}$ spacing over a $500 \times 500\,\mathrm{m}^2$ area.
Both the full-evaluation and proxy receiver grids are at a height of $1.5\,\mathrm{m}$.
We set the required spectral efficiency to $R_{\mathrm{req}}=1\,\mathrm{bps/Hz}$, the initial sample size to $N_{\mathrm{init}}=5$, and the evaluation budget to $T=100$, including the initial evaluations. \par
We compare three discretization levels for the proxy propagation database.
The \texttt{fine} setting uses azimuth and downtilt steps of $45^\circ$ and $5^\circ$, respectively, and $25\,\mathrm{m}$ receiver spacing ($N_\alpha=8$, $N_\beta=13$, $N_p=400$).
The \texttt{medium} setting uses $90^\circ$ and $10^\circ$ angular steps and $50\,\mathrm{m}$ receiver spacing ($N_\alpha=4$, $N_\beta=7$, $N_p=100$).
The \texttt{coarse} setting uses $180^\circ$ and $20^\circ$ angular steps and $100\,\mathrm{m}$ receiver spacing ($N_\alpha=2$, $N_\beta=4$, $N_p=25$).
The azimuth grids start at $0$ and exclude $2\pi$, while the downtilt grids include both $0$ and $\pi/3$.
All three settings share the same ray-tracing propagation settings and full-evaluation grid.
For the physical features, we set $\tau=0.1\,\mathrm{bps/Hz}$ and fix the input standardization statistics using $2048$ independently generated feasible configurations evaluated only by the proxy. \par
\begin{figure}[t]
    \centering
    \includegraphics[width=\linewidth]{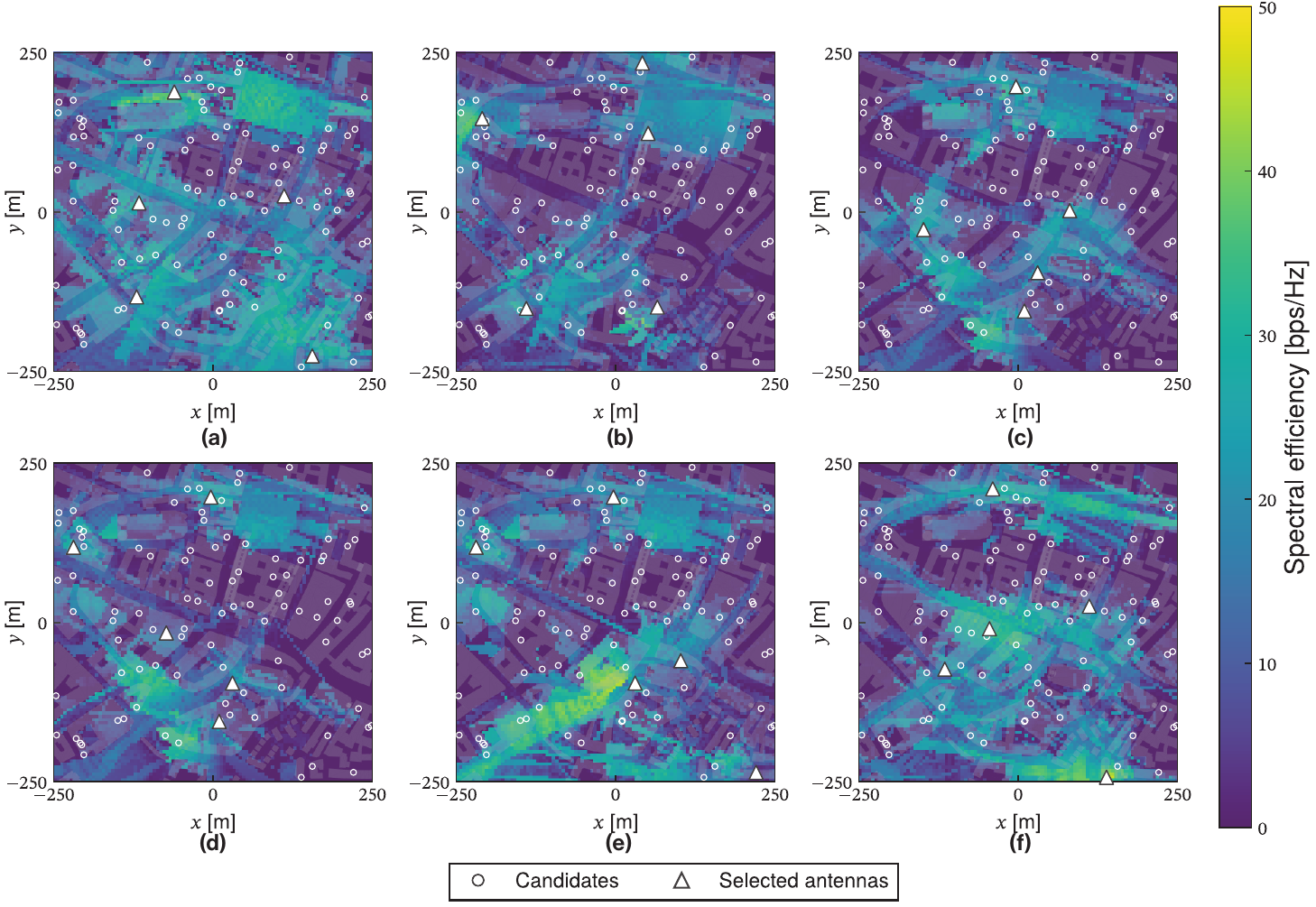}
    \caption{Optimization example in Munich: (a) Proposed (fine), (b) Random, (c) GA, (d) GP BO, (e) TuRBO, and (f) REMBO.}
    \label{fig:example_of_optimization_result}
\end{figure}
The baselines are random search (Random), a genetic algorithm (GA)\,\cite{goldbergGeneticAlgorithmsSearch1989a}, a naive GP-based BO implementation (GP BO) using a $4N = 400$-dimensional configuration representation, and the high-dimensional BO methods TuRBO\,\cite{erikssonScalableGlobalOptimization2019} and REMBO\,\cite{wangBayesianOptimizationBillion2016}.
We also include Physics BO, which uses the same three-dimensional physical features, GP, and search procedure as the proposed method but sets the prior mean to zero and learns the objective directly.
Physics BO uses the \texttt{medium} discretization in the figures, allowing comparison with Proposed (medium) to isolate the effect of the physical prior mean.
The proposed method and Physics BO use a Mat\'ern $3/2$ kernel with automatic relevance determination (ARD).
The GP targets are residuals for the proposed method and objective values for Physics BO; both are centered and scaled anew at each refit.
On the standardized target scale, the observation-noise variance is initialized to $10^{-3}$ and learned within $[10^{-4},10^{-1}]$.
Fig.\,\ref{fig:result} shows convergence performances in \texttt{Munich} ((a)(b)) and \texttt{Etoile} ((c)(d)). After 100 evaluations, Proposed (fine/medium/coarse) achieves
87.6/86.2/83.0\% coverage in \texttt{Munich} and 96.0/95.2/93.2\% in \texttt{Etoile}.
Physics BO achieves 85.5\% and 93.9\%, while TuRBO, the strongest
remaining baseline, achieves 72.4\% and 89.6\%, respectively.
Thus, Proposed (fine) improves over TuRBO by 15.2 and 6.4 percentage
points, while using the physical prior mean improves the medium setting
over Physics BO by 0.8 and 1.3 points.\par
At $300\,\mathrm{s}$, \texttt{medium} reaches 78.4\% coverage in Munich and 93.1\% in Etoile. \texttt{fine} provides the best performance for a fixed
evaluation budget but incurs a larger precomputation cost, whereas
\texttt{medium} provides a favorable balance between the two.
These results show the tradeoff between evaluation efficiency and precomputation cost when choosing the proxy discretization.

\section{Conclusion}
This paper proposed an efficient BO framework for MIMO BS design problem.
The method uses a physics-based proxy to provide both the prior mean and a three-dimensional feature representation.
A residual GP corrects the propagation-based approximation, improving the fraction of receiver locations that meet the spectral efficiency requirement within a limited budget of full evaluations.

\end{document}